\documentclass[compsoc,conference,a4paper,10pt,times]{IEEEtran}
\IEEEoverridecommandlockouts
\usepackage{cite}
\usepackage{amsmath,amssymb,amsfonts}
\usepackage{algorithmic}
\usepackage{graphicx}
\usepackage{textcomp}
\usepackage{bmpsize}
\usepackage{xcolor}
\usepackage{lipsum}
\usepackage[colorlinks=true,urlcolor=black]{hyperref}
\def\BibTeX{{\rm B\kern-.05em{\sc i\kern-.025em b}\kern-.08em
    T\kern-.1667em\lower.7ex\hbox{E}\kern-.125emX}}

\newcommand{\grantsponsor}[3]{#2 (#3)}
\newcommand{\grantnum}[2]{#2}

\begin{document}

\title{Towards Comprehensive Legislative Requirements for Cyber Physical Systems Testing in the European Union%
\thanks{This research was funded by the CyberExcellence by DigitalWallonia project (No. \grantnum{SPW}{2110186}), funded by the \grantsponsor{SPW}{Public Service of Wallonia}{SPW Recherche}.}%
}

 \author{\IEEEauthorblockN{Guillaume Nguyen}
 \IEEEauthorblockA{\textit{NADI, Comp. Sc. Faculty} \\
 \textit{Univesity of Namur}\\
 Namur, Belgium \\
 guillaume.nguyen@unamur.be}
 \and
 \IEEEauthorblockN{Manon Knockaert}
 \IEEEauthorblockA{\textit{NADI, Law Faculty} \\
 \textit{Univesity of Namur}\\
 Namur, Belgium \\
 manon.knockaert@unamur.be}
 \and
 \IEEEauthorblockN{Michael Lognoul}
 \IEEEauthorblockA{\textit{NADI, Law Faculty} \\
 \textit{University of Namur}\\
 Namur, Belgium \\
 michael.lognoul@unamur.be}
 \and
 \IEEEauthorblockN{Xavier Devroey}
 \IEEEauthorblockA{\textit{NADI, Comp. Sc. Faculty} \\
 \textit{Univesity of Namur}\\
 Namur, Belgium \\
 xavier.devroey@unamur.be}
 }

\maketitle
\thispagestyle{plain}
\pagestyle{plain}

\begin{abstract}
While procedures prevail on the European market for the greater good of its citizens, it might be daunting when trying to introduce a product, whether innovative or not. In the current world, Cyber-Physical Systems (CPSs) are ubiquitous in our daily lives. Cars can provide intrusive assistance as they can brake or turn wheels on their own, buildings are getting smarter to optimize energy consumption, smart cities are emerging to facilitate information sharing and orchestrate the response to emergency situations, etc. As the presence of such tools will grow in the coming years and people will rely even more on CPSs, we certainly need to ensure that they are safe and reliable for users or everybody else, which is why regulations are so important. However, compliance should not act as a barrier to new actors coming to the European market. Nor should it prevent current actors from keeping systems deemed compliant when introduced while obsolete at the time they are used. While the individual elements we point out might not bring novelty in the various research areas we cover (EU policies, requirements engineering, business engineering, and software engineering), this paper identifies the challenges related to building and testing a CPS with respect to applicable laws and discusses the difficulty of automating the response to those challenges, such as finding a relevant legal text, paying for mentioned materials or identifying the level of compliance to a legal text. Our analysis of the holistic context when considering the compliance testing of CPS provides an overview enabling more effective decision-making as well.

\end{abstract}


\section{Introduction}
Designing, producing, and selling a legally compliant and safe (cyber-)secure product in the European Union (EU), which satisfies privacy concerns for EU citizens and other users, is quite challenging. Indeed, products such as cars or drones (unmanned aerial vehicles) embark embedded computer systems for control purposes. Such systems directly affect the safety of such products and should be closely assessed when allowing their introduction into the European market. Indeed, Operational Technologies (OTs) are an omnipresent component of our daily lives. This type of technology is defined by Gartner as \textit{hardware and software that detects or causes a change, through the direct monitoring and/or control of industrial equipment, assets, processes, and events} \cite{gartnerDefinitionOperational}. Whether production lines of sensitive medicines with precise dosages with Industrial Control Systems, Advanced Driver Assistance Systems (ADAS) on cars, or water waste toxicity control systems would all have critical (even deadly) consequences if their system failed. OT systems composed explicitly of at least one sensor and one actuator, connected through a network to a computer (processor, microprocessor, etc.), are called Cyber-Physical Systems (CPS). CPSs share features with Embedded Systems and the Internet of Things (IoT). For example, an ADAS that detects when a car goes off the lane without having its turning signals on can make the car turn to go back into the lane. As previously said, such systems have an impact on the natural world. Thus, they can affect human environments, and the security and safety of such systems are directly linked with the security and safety of human operators, users, and bystanders. Indeed, a failure of the system to behave as designed or a failed design could lead to significant damages in the real world or even human deaths.

As we focus on European companies or companies with economic and business interests in the EU, we must consider legal texts adopted by the EU. By legal texts, we mean Regulations, Directives,\footnote{Beware that EU Directives are sets of goals that act as the basis for member states to devise their own laws.} Decisions \cite{europaTypesLegislation} as well as implementing and delegated acts \cite{europaImplementingDelegated}.\footnote{For the sake of simplicity but keep in mind that those delegated and implementing acts (or decisions) are not exactly legislative texts in strict EU law terms as they do not follow the legal and mandatory approbation process.} However, those might not offer the right level of specification required to perform technical tests. Nonetheless, some organizations offer a deeper level of detail. In a nutshell, the EU counts three European Standard Organisations (ESO) allowed to publish European Standards (EN) after a request from the EC following \textit{Regulation (EU) No 1025/2012 on European standardization} \cite{europaEURLexx}:
\begin{enumerate}
    \item the European Committee for Standardisation (CEN);
    \item the European Committee for Electrotechnical Standardisation (CENELEC);
    \item the European Telecommunication Standard Institute (ETSI).
\end{enumerate}

Standardization and Regulation are key in shaping patterns of production and consumption \cite{Murphy2009}. There is a true need for compliant product models for any type of product or service (connected or not). Nowadays, with the growing concerns in terms of cybersecurity the European regulatory and standardisation scheme level of suitability needs to be addressed. 
Kamara \cite{Kamara2024} analyzed the current scheme in relation to the Cybersecurity Resilience Act and concluded that the level of collaboration between the EU and ESO was key in the degree of positive impact in terms of cybersecurity on the European market. Kamara also noted that Cybersecurity (and privacy) was already a concern for all ESOs while they were encountering difficulties in collaborating with other organisms (while the reason was not yet academically researched). Furthermore, cybersecurity is one thing, but compliance in general (safety, etc.) is also crucial in producing, selling, and using goods on the European market. Thus, with the rethinking of the collaboration between the EC and ESOs, there is a genuine opportunity to include more straightforward compliance issues (technical or not) and software-related concerns for every product and service produced, sold, and used on the European market.

However, the standards related to specific legal texts might not be sufficient to produce technical compliance tests for Cyber-Physical Systems. In this paper, we devise a method to find the lowest level of specifications while keeping the traceability to a legal text to help engineers perform those tests. For this purpose, we analyze relevant regulations, directives, standards, documents, tools, etc., created and/or referenced by official sources in the EU for each application domain laid out by Tekinerdogan et al. \cite{Tekinerdogan2020-sj} in the context of CPSs. The contributions of this paper are:
\begin{enumerate}
    \item A method for identifying a system's test requirements for legal compliance
    \item An empirical link between European legal texts and industrial CPS for each domain of application.
    \item A critic of the current compliance process concerning CPS in the EU or ``Challenges''.
    \item An approach towards automatized legal compliance checking using available technologies and current research.
\end{enumerate}

\section{Background} \label{sec:Background}
In this section, we provide a deeper context concerning CPS, standardization in the EU, and legal compliance in the EU, with a note on product compliance. We point out relevant concerns when automating the compliance testing of CPS in the EU.

\subsection{Cyber-Physical Systems} \label{sec:Back-CPS}
As previously introduced, multiple concepts, such as OT, Embedded Systems, and the Internet of Things, seem to collide when discussing CPS. An early example of CPS could be the \textit{Apollo Guidance Computer} (AGC), which was embedded in the rocket and the lunar module. It was used to operate both to travel through space and land on the moon. However, a human operator (i.e., Neil Armstrong) had to guide the lunar module during the landing, as scientists programming the computer could not have anticipated the type of surface the module would have to land on \cite{Hall1996}. Based on such systems, the first \textit{Digital Fly-By-Wire} (DFBW) system was introduced on an aircraft in 1972 and was used to interpret the motion of the pilot's control stick to compute the required movement of each surface of the plane \cite{Jenkins_2001}.

The term \textit{Embedded Computer Systems} (ECS) probably came from L/C Manley in 1975 \cite{manley1975} in the \textit{Final Report of the Joint Logistics Commanders Electronic Systems Reliability Workshop} \cite{united1975final}. This paper talks about software reliability and how ECS needed to benefit from better software quality alongside related intellectual challenges like real-time computing, process control, reliability, etc. They mainly proposed policy and procedural changes to enhance the reliability and overall quality of software used for ECS. This workshop most probably triggered the US DoD 5000.29 directive on the \textit{Management of Computer Resources in Major Defense Systems} (26 April 1976), after which ADA programming language was chosen for ECS \cite{downes1982}.

The \textit{Internet of Things} (IoT) is believed to take its roots in the US DoD ARPANET project (1969), basically aiming at interconnecting various computing resources for the sake of productivity and efficiency \cite{bolt1981}. Later in the 1990s, Mark Weiser came up with the concept of \textit{Ubiquitous Computing} or more recently named \textit{Pervasive Computing} or \textit{Ambient Intelligence} and refers to the invisible presence of computers in many objects, allowing users to access and exchange data no matter their positioning on the globe \cite{Weiser1999}. Afterward, \textit{Sensor Nodes} combined with technologies such as \textit{Wireless Communication} and \textit{Digital Electronics} allowed for the emergence of wider sensor networks \cite{Akyildiz2002}. Alongside technologies such as \textit{Big Data}, \textit{Cloud Computing and Storage}, and \textit{Nanotechnologies}, IoT is continuously evolving \cite{IbarraEsquer2017}.

Cyber-Physical Systems (CPS) first appeared in the literature in 2006 during a workshop by the National Science Foundation (NSF) of the United States. This workshop led to the creation of three founding papers from Lee et al. \cite{Lee2007}, Sha et al. \cite{Sha2008-ot}, and Rajkumar et al. \cite{Rajkumar2010-jj}. The definition that we retain is the following:
\vspace{12pt}
\begin{quote}
    ``\textit{Cyber-physical systems (CPS) are physical and engineered systems whose operations are monitored, coordinated, controlled, and integrated by a computing and communication core. This intimate coupling between the cyber and physical will manifest from the nano-world to large-scale wide-area systems. The internet transformed how humans interact and communicate with one another, revolutionized how and where information is accessed, and even changed how people buy and sell products. Similarly, CPS will transform how humans interact with and control the physical world around us.}'' \cite[p.~1]{Rajkumar2010-jj}
\end{quote}
\vspace{12pt}

Furthermore, IoT is a ``\textit{concept used to define or reference systems that rely on autonomous communication of a group of physical objects}'' \cite{Espada2014}. We can see that IoT and CPS (and embedded systems) are linked and should be considered together when defining and classifying CPSs. In a nutshell, one could consider that embedded systems are CPS components that use IoT to communicate. When bringing CPS and IoT together, Liu et al. suggest that CPS ``\textit{... deeply integrates the ability of computing, communication, and control based on information acquisition in IOT.}'' \cite[p.~28]{Liu2017-qo}.

More recently, Lee et al. defined CPS as an approach to embedded systems. They define IoT as a means to interconnect sensors and actuators through networks to an interface inspired by the IT world, such as \textit{Web Interfaces}. However, while IoT could fit the earlier definition of CPS, it is unsuitable for time-critical real-world interactions. Indeed, \textit{Real-Time Control} and \textit{Safety-Critical Systems} require low-level logic and architectural designs. Lee et al. intend to give an introductory course on all the technical challenges when designing a CPS \cite{Lee2016-ql}.

\begin{figure*}
    \centering
    \includegraphics[width=0.7 \textwidth]{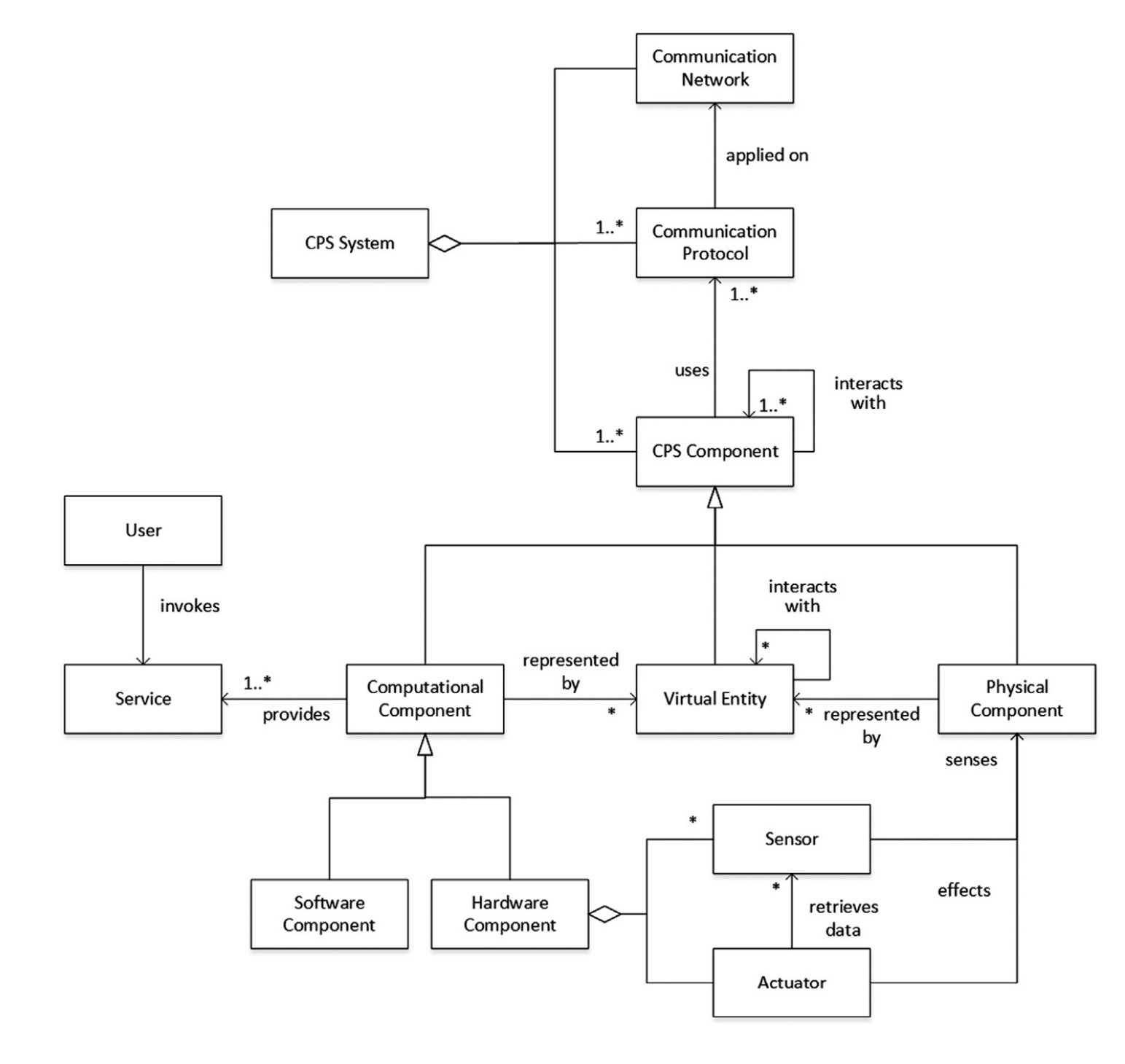}
    \caption{The metamodel by Tekinerdogan et al. \cite[p.~47]{Tekinerdogan2020-sj}}
    \label{fig:TekinerdoganMetamodel}
\end{figure*}

When focusing on identifying a CPS, the University of Berkeley provides us with a concept map around CPS. This map describes a CPS as a feedback system that requires cyber-security, safety, improved design tools, and design methodology and has an application in a specific domain \cite{CPSMapBerkeley}. Pushing further the description of a CPS, Tekinerdogan et al. \cite{Tekinerdogan2020-sj} provide us with a metamodel (Figure \ref{fig:TekinerdoganMetamodel}) and a feature-based ontology.
They describe a CPS as having constituent elements, non-functional requirements, an application domain, a discipline (software engineering, civil engineering, etc.), and an architecture. The constituent elements of a CPS are the following \cite{Tekinerdogan2020-sj}:
\begin{itemize}
    \item \textbf{Cyber} - element controlling and communicating with the other elements
    \item \textbf{Control} - behaviour modification through feedback loops. The most important component for a CPS core functionality 
    \item \textbf{Human} (non-mandatory) - Humans playing a role in achieving the modification of a behavior
    \item \textbf{Network} - configuration and communication
    \item \textbf{Physical} - sensors, actuators, plant, controller, and environment
\end{itemize}

The ten application domains are Health, Smart Manufacturing, Transportation, Process Control, Defence, Building Automation, Robotic Services, Critical Infrastructure, Emergency Response, and Other.

\begin{figure}
    \includegraphics[width=0.48 \textwidth]{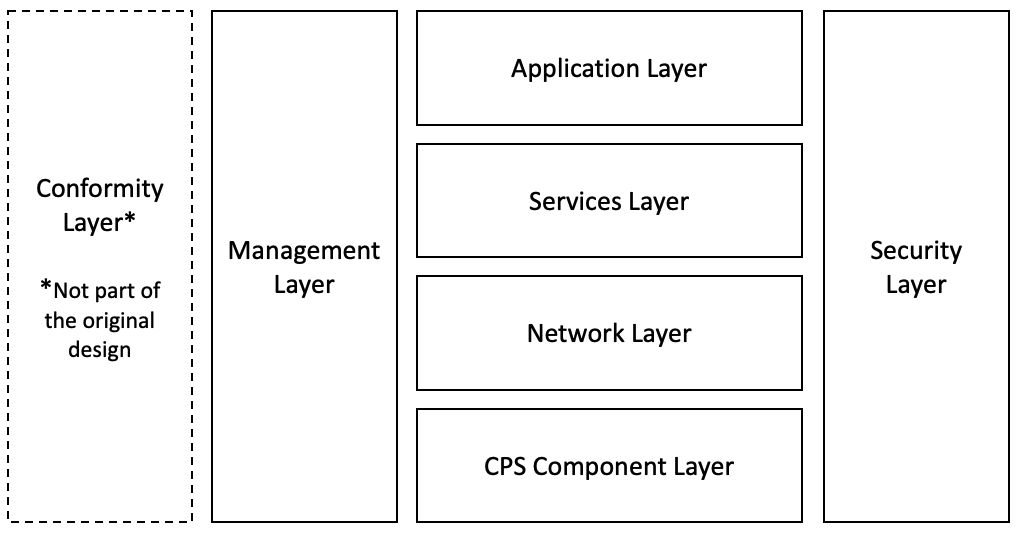}
    \caption{The CPS architecture Tekinerdogan et al. \cite[p.~59]{Tekinerdogan2020-sj}. The dashed "Conformity Layer" is a contribution of this paper.}
    \label{fig:TekinerdoganArchitecture}
\end{figure}

There are many non-functional requirements regarding CPS, such as reliability, availability, safety, and security. The high criticality of such systems requires defining many such non-functional requirements before producing and using CPS. Concerning the architecture of a CPS, Tekinerdogan et al. define it as containing four layers and two main cross-cutting concerns, as shown in Figure \ref{fig:TekinerdoganArchitecture}. We can see that the concept of CPS encloses quite various areas of concern. Indeed, software, hardware, and human components must be considered when performing compliance testing.

\subsection{EU Standardization Landscape}
\begin{figure}[t]
    \includegraphics[width=0.5 \textwidth]{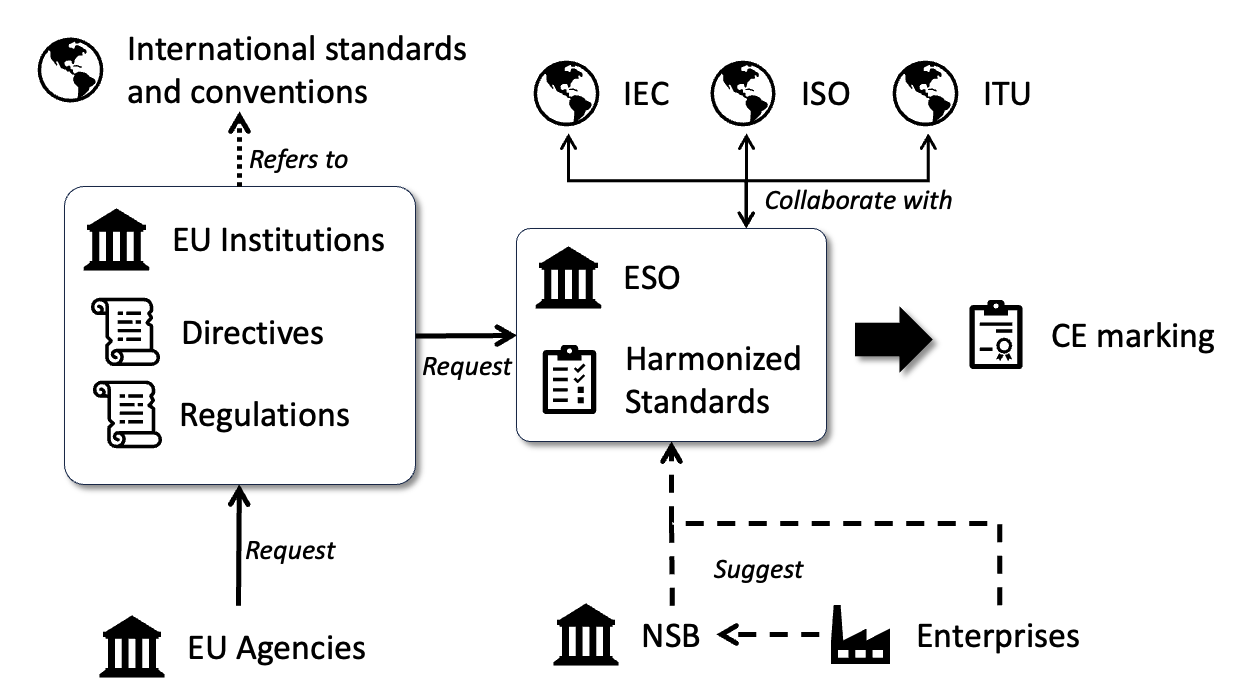}
    \caption{The reduced and simplified EU legal texts and standards scheme}
    \label{fig:three}
\end{figure}

Figure \ref{fig:three} gives an overview of the EU regulatory and standardization scheme. It gives out the path from a (EU) legal text to a European conformity marking (CE). A CE assessment is required for some categories of products sold or used in the EU. Such assessment includes a thorough procedure to be followed by manufacturers. Since 2008, the EU set out the \textit{New legislative framework}. Currently, multiple texts are supporting the market surveillance of products including those three:
\begin{itemize}
    \item Regulation (EC) 765/2008 setting out the requirements for accreditation and market surveillance relating to the marketing of products \cite{europaEURLexx}
    \item Decision 768/2008 on a common framework for the marketing of products \cite{europaEURLexx}
    \item Regulation (EU) 2019/1020 on market surveillance and compliance of products \cite{europaEURLexx}
\end{itemize}
To guide us through this new legislative framework, the EC shared \textit{The ‘Blue Guide’ on the implementation of EU product rules 2022} to help actors understand what is expected of them \cite{ECBlueGuide}. Showing the continuous involvement of the EC

The CE marking can be obtained by implementing harmonized European Standards (hEN), a simplified procedure set up by the EU. An hEN is a specific collection of European standards produced by ESO and requested by the EC \cite{europaHarmonisedStandards} that can also include standards made by the International Standard Organisation (ISO), the International Electrotechnical Commission (IEC) and the International Telecommunication Union (ITU) as set out by \textit{Regulation (EU) No 1025/2012 on European standardization} \cite{europaEURLexx}. Those harmonization requests are often composed of a collection of existing European Standards (EN), which can or should be modified by the relevant ESO when necessary, and new standards should be created. Concerning European standardization actors, the National Standardisation Bodies (NSB) are responsible for developing European consensus for the EU member states. Similar entities exist for small and medium companies, consumers, workers, and public authorities. Other organizations might get involved in this scheme via the NSB for CEN- and CENELEC-related standards and with ETSI directly for telecommunication standards by emitting standardization requests as well \cite{europaPlayersEuropean}. To summarise, many actors might make the European standardization scheme hard to follow and orchestrate while remaining necessary to understand everybody's requirements.

Furthermore, Figure \ref{fig:three} does not consider the possibility of getting a CE marking for a product based on other standards or proprietary internal procedures. Indeed, those case-specific applications and their compliance with relevant legal texts must be thoroughly documented for the assessment body to perform a full review and decide whether to grant the CE marking. On their website, the EC lays out all the steps to follow to get the CE marking and the categories of products that require such marking \cite{europaManufacturers}.

\subsection{Legal compliance in the EU}
The EU policy landscape is quite complex. There are multiple levels to take into account when talking about compliance and enforcement. Let's recall that EU Directives are expected to be implemented in the member states' national laws. Furthermore, we need to differentiate the practical compliance of the implementation of an EU Directive in a national law and the practical compliance of a company within a member state to the national law based on an EU Directive. Batory showed that the implementation of Directives could diverge from one member state to another leading to litigious cases (e.g., treaty violations) between member states where the EC has to take actions \cite{BATORY2016}. 
Differentiated integration \cite{Holzinger2012} is also a challenge for countries expected to comply with EU policies. Indeed, Zhelyazkova showed that countries with selective implementation or non-implementation of EU policies lead to higher costs when such countries are required to comply to continue their activities with the EU \cite{Zhelyazkova2014}. 
Thomann and Sager showed that the level of practical compliance of national implementers is directly linked with the quality of EU Directives. They also showed that the tools allowing the EU to check practical conformance were better than expected but insufficient. Furthermore, the incentives and deterrents related to the EU Directive compliance process need to be better understood \cite{Thomann2017}. 
Zhelyaskova, Kaya, and Schrama analyzed the decoupling between practical (field compliance to a Directive) and legal (transposition of an EU Directive into the national law) compliance between the 27 member states based on fewer policy areas. They showed that legal compliance did not always mean practical compliance and that this decoupling is more frequent when the member state's institutional capacities or the societal perception of the legitimacy of EU Directives is lower. However, the policy preferences of political actors did not seem to have a considerable impact on practical compliance \cite{Zhelyazkova2016}. 
However, practical implementation could exceed the legal compliance with an EU Directive. Indeed, the same authors showed that national actors exceed expectations when the national laws are ambiguous or incomplete and that positive societal perception of the legitimacy of EU Directives had a positive impact \cite{Zhelyazkova2017}. 
Zhelyazkova and Schrama also add that the EC may rely on expert reports to act on a member state's failure to comply with an EU Directive. Indeed, they show that a failure to adhere practically (as shown by expert reports) would likely trigger enforcement actions from the EC. At the same time, legal compliance (from EU Directive to national laws) would not specifically have the same effect. It would seem that the EC prioritizes practical over legal compliance. Following previous results, they also found that (pro-EU) civic participation was an enabler of practical compliance across member states \cite{Zhelyazkova2023}.

However, those empirical reports cover the application of EU Directives within four areas (i.e., Internal Market, Justice and Home Affairs, Environment, and Social Policy). In the context of CPS and products in general, one must assess the practical implementation from various industrial actors across industry domains. As mentioned above, the CE marking supposedly requires such assessment while missing out on the actual implementation (i.e., the actual end-product) by only relying on the documentation provided by the manufacturer related to the production of the to-be \textit{marked} product. Indeed, in section \textit{5.1.3. Actors in conformity assessment – Positioning of conformity assessment in the supply chain} of the \textit{Blue Guide}, we can see that while the legislator provides the legislative requirements, the manufacturer can decide (when no other legislative requirements) to carry the product assessment itself or by requesting the help of a third party (with respect to section \textit{5.2.3. Competence of notified bodies}) \cite{ECBlueGuide}. This means that potentially, no EU-compliant notified bodies would assess a product's compliance before its introduction to the market. However, in some cases a specific legal text might exclude this possibility when assessing for the CE marking. Indeed, the recently adopted \textit{Regulation of the European parliament and of the council on horizontal cybersecurity requirements for products with digital elements (Cyber Resilience Act)}\footnote{Adopted by the council on October 10th 2024} \cite{europaEURLexx} includes specific demands for Class II systems. Those systems include, among other things, IoT, Industrial Automation \& Control Systems, sensors and actuators. And requires the involvement of an EU notified body \footnote{cf. Module H in Annex VI} or following an EU-type examination procedure \footnote{cf. Module B and C in Annex VI}

Nevertheless, the manufacturer is left with full responsibility in case of failure to comply with legislative requirements (when no other legislative requirements are in force). Fortunately, the \textit{Market Surveillance} mechanism requires member states to appoint one or multiple \textit{Market Surveillance Authorities} tasked with ensuring the conformity of products already introduced on the market. Each member state has its own strategy concerning market surveillance. However, non-conformity can be observed and reported by any concerned party to market surveillance authorities, which notifies the related economic actor. After a short period of time, if the non-conformance is recognized, actions can be taken to withdraw the product from the market. Of course, Regulation (EU) 2019/1020 \cite{europaEURLexx} adds that non-conformance should be detected before introducing a product to a market, but non-compliant products seem to reach markets anyway. A list of non-compliant products can be found on the \textit{Safety Gate} platform of the EC \footnote{https://ec.europa.eu/safety-gate-alerts/screen/webReport}. Following the Cyber Resilience Act, they suggest the use of the number of software vulnerabilities found after the go-to market to monitor the practical readiness of software after testing and assessment.

Concerning sanctions in case of non-compliance, the member states are trusted with the magnitude of the sanctions. Furthermore, some articles within EU Legal Texts, such as articles 83 and 84 of EU Regulation 2016/0679 (GDPR) foresee ``\textit{administrative fines up to 20 000 000 EUR, or in the case of an undertaking, up to 4 \% of the total worldwide annual turnover of the preceding financial year, whichever is higher.}''\cite[p.~82-83]{europaEURLex32016R0679}. Similarly, articles 34, 35, and 36 of EU Directive 2022/2555 (NIS2) foresee ``\textit{administrative fines of a maximum of at least EUR 10 000 000 or of a maximum of at least 2\% of the total worldwide annual turnover in the preceding financial year of the undertaking to which the essential entity belongs, whichever is higher.}'' \cite[p.~60-61]{europaEURLex32022L2555}.

\section{Related Work} \label{sec:RelatedWork}
We talked about CPS and EU-specific context, and we showed that legislative requirements were left to the free assessment by companies looking to put the CE marking on their products while guided by EN or hEN when available and even external help from \textit{Notified Bodies}. However, interpreting, let alone testing, legislative requirements or international standards remains challenging. This section presents related work in the context of testing text-based requirements.

Concerning Information Security regulations for CPS, Urquhart et al. offered a quick analysis related to the NIS Directive and GDPR Regulation \cite{urquhart2017assessing}. They show how IT concerns transpose into the OT world. However, those directives mainly focus on information security (and privacy), while CPSs also have safety concerns that may or may not be regulated. Concerning domain-specific CPS, Shenoy et al. proposed an analysis of the regulatory landscape for Unmanned Aircraft Systems (UAS or drones). They showed how privacy concerns should be adapted to follow the fast-evolving drone technology (i.e., cargo-carrying drones, etc.) \cite{Shenoy2022-fk}. 

Furthermore, work has already been performed in Requirements Engineering (RE) and legal regulations. Sienna et al. and Ingolfo et al. worked on a RE Framework called \textit{Nomos} \cite{Siena2009} \cite{Ingolfo2014}. The framework is centered on rights that actors have and must respect, considering specific actions (or context). The rights are characterized following the Hohfeldian taxonomy \cite{Hohfeld1913} and by joining correlated rights. For example, the right of an individual to access a resource (i.e., a \textit{claim}) means the \textit{duty} of another individual to provide such a resource. Such relation will be referred to as \textit{ClaimDuty}. The framework includes requirements concerns related to the law and satisfyingly captures the relationship between both. Breaux et al. developed a method to extract privacy and security-specific requirements from regulatory rules \cite{Breaux2008}. They create a catalog with the various constraint types to classify the rules then they deal with the exceptions and priorities. This produces a model of priorities between the multiple rules that can be used to check the conformance to a specific regulatory text. 

On the other hand, Kiyavitskaya et al. worked on a way to automatize the extraction of \textit{Rights} and \textit{Obligations} \cite{Kiyavitskaya2008}. They developed the Cerno framework to automatically parse legal documents using grammar, annotate with domain-specific tags, and map the rights and obligations to a database. Their method performs well but is limited in the presence of  \textit{Constraints}. On a more technical side, Winter et al. use Natural Language Processing (NLP) techniques on regulatory documents to extract constraints and their relations \cite{Winter2018}. Their method consists of three steps: (1) Pre-Processing, where they slice the documents into sentences then (2) Processing to group sentences based on the subject (matter); and finally, (3) Post-Processing produces a graph showing the various relations within the analyzed text. Their method can work on any documents where requirements can be extracted (legal texts, standards, etc.). 

Interestingly, there is extensive research on automating compliance with the General Data Protection Regulation (GDPR). Aberkane et al. performed a systematic mapping study on the matter in 2021, and we can see that such concerns bring NLP and RE together \cite{Aberkane2021}. Out of 448 relevant papers, they found that papers were mainly focused on NLP and RE (420), while nine papers linked GDPR and NLP and 20 papers for GDPR and RE. Thus it seems that using NLP techniques for RE appears to be a popular method even in the context of Legal Texts such as the GDPR.

Concerning the interpretation of legal text using Machine Learning, Sleimi et al. developed a model to extract legal metadata from legal texts \cite{Sleimi2021}. Among other things, it allows a user to distinguish mandatory and prohibited actions in the various articles. One could add another layer using a Large Language Model (LLM) \cite{Weizenbaum1966} to distinguish CPS features from articles, sub-articles, etc. Then, we would need a way to go from those features to testable specifications. However, the quality of the results and the accountability in case the LLM delivers false or incomplete information still need to be assessed. 

Torre et al. designed a method of compliance checking based on UML modeling for GDPR \cite{Torre2019}. Their approach is composed of four steps: (1) Building the conceptual model, (2) Tailoring it, (3) Generating an instance of the domain model including contextual information such as relevant national laws of a member state, and finally (4) Checking compliance. However, they did not complete steps (3) and (4) that were left for future work. Similarly, Sacre et al. also addressed this issue by using models \cite{Sacre2021}. Their approach implied extensive manual work in understanding the legal and product requirements and building a meta-model of a compliant implementation for a specific product. Then, to assess the compliance of the actual product, they compare the hypothetical but compliant meta-model to the actual one and list discrepancies between the two. A compliant product would have no critical discrepancies.

We can also find papers focusing on the GDPR when looking at OT-related domains. Kaneen et al. use a model-based approach to the GDPR for IoT \cite{Kaneen2020}. Their methodology includes a UML class diagram for each of the ten questions they elaborated based on GDPR (e.g., \textit{Where is the data stored?}). They did not include all the GDPR concerns, and their methodology implies efforts at an initial phase of a system design; however, they successfully proposed a working GDPR testing methodology. Considering industrial IoT, Demertzi et al. suggest a framework enabling the privacy-safe feeding of data-based defect detection machine-learning algorithms using federated learning to decentralize and distribute data processing and blockchain to ensure integrity and secure data sharing \cite{Demertzi2023}.

\begin{figure}[t]
    \includegraphics[width=0.48 \textwidth]{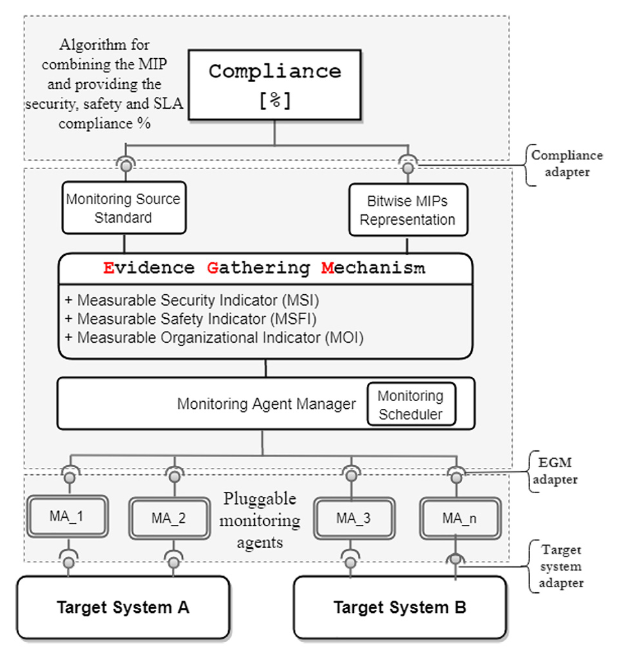}
    \caption{The Monitoring and Standard Compliance Verification (MSCV) framework from Bicaku et al. \cite{Bicaku2020}}
    \label{fig:five}
\end{figure}

Concerning standard compliance checking, Bicaku et al. show the importance of standardization in CPS \cite{Bicaku2019} and industrial IoT \cite{Bicaku2020}. Their work focuses on automating industrial IoT and CPS application compliance checks. They identified relevant compliance frameworks, European projects, and scientific research. Figure \ref{fig:five} shows the Monitoring and Standard Compliance Verification (MSCV) framework they developed. By extracting Measurable Indicator Points (MIPs), which include security, safety, and/or organizational concerns, they set up a framework allowing for the system to check for the presence of indicators related to a specific standard. By connecting monitoring agents to the target systems and using an evidence-gathering mechanism, they managed to compute the level of compliance with said standards with respect to the tested requirements. A MIP is defined by
\begin{enumerate}
    \item a Source (A specific standard),
    \item a Definition,
    \item a Monitoring Plugin (Explaining how the MIP conformance will be assessed) and
    \item a Monitoring Value (usually True or False).
\end{enumerate}

As we can see, encouraging work has been performed across various research areas. However, the link between legal text compliance and CPS testing is not quite established, while it is undoubtedly pre-paved. Yet, a way between requirements from legal texts or standards and automatic conformity testing for CPS seems realizable while still necessitating tremendous (but one-time only) manual work at earlier process steps.

\section{Test requirements} \label{sec:TestRequirements}
In the previous sections, we provided a bit of context on the technologies involved, how they interacted, and what could be expected of EU legal texts. In this section, we discuss the interaction between the various testing levels of CPS and what could be expected and/or required from legal texts. First, to keep consistency with research the work performed by Tekinerdogan et al. \cite{Tekinerdogan2020-sj} we will focus on the components presented in section \ref{sec:Back-CPS}. As CPS relates to various components, the testing levels are not limited to Spillner and Linz's \cite{Spillner2011-il}:
\begin{itemize}
    \item ``\textbf{Component Testing} involves systematically checking the lowest-level components in a system's architecture. Depending on the programming language used to create them, these components have various names, such as \textit{units}, \textit{modules}, or, in the case of object-oriented programming \textit{classes}. The corresponding tests are therefore called \textit{module tests}, \textit{unit tests}, or \textit{class tests}.'' \cite[p.~58]{Spillner2011-il}
    \item ``\textbf{Integration Testing} is the next level following component testing. Integration testing assumes that the test objects handed over to this level are already component tested and that any component-internal defects have been corrected as far as possible.'' \cite[p.~66]{Spillner2011-il}
    \item ``\textbf{System Testing} is the next level once integration testing is complete. This level checks that the complete, integrated system fulfills its specified requirements.'' \cite[p.~74]{Spillner2011-il}
    \item ``\textbf{Acceptance Testing} takes place from the customer/end-user's viewpoint and is the only test that the customer is directly involved in or is actually responsible for.'' \cite[p.~76]{Spillner2011-il}
\end{itemize}
Indeed, for each level, the \textit{Physical} component should be taken into account as hardware plays a critical role in CPS, whether it can sustain what is asked of it (processing, etc.) or the environment where it should run (low temperature, high humidity, etc.). The \textit{Network} component should also be added as it is used for communication between the various components. Then the \textit{Human} component should be offered proper training before putting a CPS to the test however, this relates to \textit{Acceptance Testing} while training material should be offered by the manufacturer. 

When considering testing levels for CPS, Abbaspour et al. provide us with a survey on the various techniques available based on various research papers where they determined six testing levels. Besides \textit{Integration Testing} and \textit{System Testing} we have \cite[p.~197]{AbbaspourAsadollah2015}: 
\begin{itemize}
    \item ``\textbf{Hardware Testing} consists of testing hardware components of CPS, including tests of each component's functionality based on the system requirements'' \cite[p.~197]{AbbaspourAsadollah2015}
    \item ``\textbf{Structural and Computation Testing} is normally based on the detailed design and not on the required functions of the program. However, for computation testing, the tester (or developer) uses the program's structure and chooses paths that are used to recognize domains.'' \cite[p.~197]{AbbaspourAsadollah2015}
    \item ``\textbf{Extra-Functional Properties Testing (EFP Testing)} Multiple parts of a CPS are often embedded systems executing real-time software. For such a system, it is required to guarantee both (1) function correctness and (2) non-functional or extra-functional correctness. Extra-functional properties are closely related to the inherent interaction with the system environment. Examples of such properties may be temperature, power consumption, and timing'' \cite{Sha2008-ot} \cite[p.~197]{AbbaspourAsadollah2015}
    \item ``\textbf{Network Testing} the tester is assisted for verifying, controlling or comparing the performance by recording the current state of network operation.'' \cite[p.~198]{AbbaspourAsadollah2015}
\end{itemize}

Those testing levels are necessary to perform holistic testing on CPS (i.e, each level should be covered). Furthermore, as we established the link between CPS and IoT, we could also consider the following levels as presented by Dedonno et al. \cite{DeDonno2019}: 
\begin{itemize}
    \item ``\textbf{Edge computing} is being promoted as a strategy to achieve scalable and highly available Web services. It pushes business logic and data processing from corporate data centers out to proxy servers at the “edge” of the network'' \cite[p.~1]{Pang2004}
    \item ``\textbf{Fog Computing} is a highly virtualized platform that provides compute, storage, and networking services between end devices and traditional Cloud Computing Data Centers, typically, but not exclusively located at the edge of the network.'' \cite[p.~13]{Bonomi2012}
    \item ``\textbf{Cloud computing} is a model for providing or obtaining information and communication technology (ICT) services over a network like the internet. The services provided may include servers, operating systems, networks, software, storage, and applications. Cloud services are provided through a scalable and 'elastic' pool of shareable resources based on physical resources (hardware) in purpose-built data centers. Using self-service provisioning, configuration and administration, users can rapidly ramp up (or down) their ICT capacity as demand changes.'' \cite[p.~3]{EP2016}
\end{itemize}

As Rajkumar et al. \cite{Rajkumar2010-jj} pointed out, with so many different testing levels, it might be challenging to perform holistic testing on CPS with a single method. However, encouraging testing methods such as the CISRT middleware from Huang et al. \cite{Huang2010} cover all the testing levels from Abbaspour et al. \cite{AbbaspourAsadollah2015}. Furthermore, when testing non-functional requirements, other challenges may arise. Indeed, when considering \textit{Security}, Pfleeger et al. \cite{Pfleeger2010} and Jansen et al. \cite{jansen2010directions} showed that it might not be possible to test all possible security requirements due to the complexity, variety, and even interaction of security issues. Security testing requires a non-negligible amount of brainstorming to anticipate security issues.

\section{Application domains and related EU Legal Texts}
\begin{table*}[!htb]
    \centering
    \caption{Industry application and (EU) Legal text examples for each application domain}
    \label{tab:reg_stan_tab1}
    \begin{footnotesize}
    \begin{tabular}{|p{1.8cm}|p{5.0cm}|p{5.0cm}|p{1.4cm}|p{1.2cm}|}
    \hline
        \textbf{Domain} & \textbf{Example in the industry} & \textbf{Legal text (EU)} & \textbf{hEN} & \textbf{Standards} \\ \hline
        \textbf{Health} & Wireless medical devices presence is growing in hospital and operating rooms \cite{Mahfouz2012-as} & Regulation (EU) 2017/745 on medical devices \cite{europaEURLex32017R0745} \& Regulation (EU) 2016/746 on in vitro diagnostic medical devices \cite{europaEURLex32017R0746} & Mandates 565, 575 \& 575 Amd 1 & Yes \\ \hline
        \textbf{Smart\newline Manufacturing} & Smart factories through industry 4.0 aim at increasing the efficiency of the product line either by reducing the costs or improving the flexibility of the resources by using interconnected devices, sensors and actuators \cite{Lasi2014-ml} & Regulation (EU) 2023/1230 on machinery \cite{europaEURLexx} & ~ & ~ \\ \hline
        \textbf{Transportation} & Advanced driver assistance systems (ADAS) alongside with other technologies aiming at self-driving and connected cars \cite{Arena2020-bf} Multiple Assistance Systems for Civil Aviation are used for Conflict Detection and Resolution \cite{Tang2019} & Regulation (EU) 2019/2144 on type-approval requirements for motor vehicles and their trailers, and systems, components and separate technical units intended for such vehicles, as regards their general safety and the protection of vehicle occupants and vulnerable road users \cite{europaEURLexx} \& Regulation (EU) 2018/1139 of the European Parliament and of the Council on common rules in the field of civil aviation and establishing a European Union Aviation Safety Agency \cite{europaEURLexx} & Mandates 578, 576, 570, 557, 546, 546 Amd 1, 591, 598 & Yes \\ \hline
        \textbf{Process\newline Control} & Detection of chemical compounds for water pollution detection and communication with waste-water plants \cite{Garrido-Momparler2022-ks} & Regulation (EU) 2019/1009 laying down rules on the making available on the market of EU fertilizing products \cite{europaEURLexx} & Mandate 564  & Yes \\ \hline
        \textbf{Defence} & Upcoming of Lethal Autonomous Weapon Systems (LAWS) \cite{Scharre_2019} & Regulation (EU) 2021/697 establishing the European Defence Fund \cite{europaEURLexx} & ~ & ~ \\ \hline
        \textbf{Building\newline Automation} & Connected devices to help monitor and control a house and support the residents \cite{Marikyan2019-ji} & Directive (EU) 2018/844 on the energy performance of buildings \cite{europaEURLexx} & Mandates 549, 597, 599 \& 480 (no longer active) & Yes \\ \hline
        \textbf{Robotic\newline Services} & Unmanned aerial or ground vehicles (UAV, UGV) combined with other wireless sensors and data analysis technologies are used to increase the productivity of farming exploitations \cite{Moysiadis2021-ad} & Commission Delegated Regulation (EU) 2019/945 on unmanned aircraft systems and on third-country operators of unmanned aircraft systems \cite{europaEURLexx} & Mandate 567 & Yes \\ \hline
        \textbf{Critical\newline Infrastructure} & Transitioning from classical grid management tools to smarter ones to improve the efficiency of a grid \cite{Fang2012-xk} & Directive (EU) 2019/944 on common rules for the internal market for electricity  \cite{europaEURLexx} & Mandate 490 (no longer active) & Yes \\ \hline
        \textbf{Emergency\newline Response} & Using human-computer interaction (HCI) and available surveillance system to increase the security within a smart city \cite{Kashef2021-av} & Regulation (EU) 2021/836 on a Union Civil Protection Mechanism \cite{europaEURLexx} & ~ & ~ \\ \hline
        \textbf{Other} & Ingenuity mars copter had to adapt itself to the aerodynamics of Mars to perform the first successful flight there \cite{Tzanetos2022-xn} & ~ & ~ & ~ \\ \hline
    \end{tabular}
    \end{footnotesize}
\end{table*}

The next step for CPS legal compliance testing is to link existing legal texts and related materials to CPS with the various \textit{Domains of Industry} as proposed by Tekinerdogan et al. \cite{Tekinerdogan2020-sj}. In table \ref{tab:reg_stan_tab1}, we gathered industrial examples for each application domain from the literature. We searched (not exhaustively) the Official Journal of the European Union \cite{europaEURLexx} for relevant legal texts related to the industrial examples or the domain. Then, we used the eNorm platform \cite{europaENormPlatform} to search for mandates related to the example or the domain. A \textit{Yes} in the \textit{Standards} column means that we found at least one reference to a standard or convention after reading the selected legal text.

Even without a complete and full view of legal texts related to each CPS application domain, we can see that it is possible to find a relationship between both concepts. We found that \textit{Transportation} was mainly well exploited (automobiles, trains, planes, boats, etc.) and had relation with other domains such as \textit{Emergency Response} (for the distress system of boats) or \textit{Robotic Services} (for UAV). \textit{Process Control} was mainly related to quality control. \textit{Critical Infrastructure} was mainly well regulated with links to \textit{Building Automation}, which mostly had energetic efficiency as the primary concern. \textit{Smart Manufacturing} was relatively new in the legislation landscape, and we only found a unique legal text (cf. Table \ref{tab:reg_stan_tab1}). The \textit{Defence} domain had a recent regulation on the funding and ethics of European Defence. It might be a sign that military CPS manufacturers might come across limitations in developing and producing intelligent weaponry systems in the future (hopefully). During our research, we greatly appreciated how regulations on medical devices were set out. Caroll et al. suggested such an arrangement based on a thorough review of existing rules and standards at the time \cite{Carroll2016}. Their research supposedly led to the creation of two new regulations \cite{europaEURLex32017R0746} \cite{europaEURLex32017R0745} replacing the three previous ones. 

Other interesting legal texts might be relevant in the current context. While they are related to data protection and cybersecurity, which are concepts that are only starting to appear in the OT world, they will drive related (future) legal texts in other interesting directions than the more classic operational concerns: the GDPR \cite{europaEURLexx} and NIS 2 Directives \cite{europaEURLexx}. For example, medical device manufacturers are required to consider such concerns as they collect, process, and transfer patients' medical data.

By following official sources of information from the EU, we made connections to industrial concepts. International standards and conventions, as well as hEN and EN, might offer deeper levels of specifications that would greatly help us understand what is expected from manufacturers when making, maintaining, and selling related products. To see how far we can get starting from a legal text, we need to dig deeper into the material offered and referred to by the EU.

References to software components within product-related legal texts might not be as straightforward as expected, indeed we need to understand how CPS products work and which parts of those systems are related to specific legislative requirements. Concerning \textit{Transport}, Regulation (EU) 2019/2144 \cite{europaEURLexx} also known as the General Safety Regulation 2 (GSR2) includes concerns about ADAS for all types of road vehicles.
For \textit{Robotic Services} such as UAV, Implementing Regulation (EU) 2019/947 on the rules and procedures for operating unmanned aircraft \cite{europaEURLexx} includes the requirement for a defined \textit{Geographical Zone} limiting drone access to certain areas to address various risks such as privacy or environmental security. For \textit{Smart Manufacturing}, Regulation (EU) 2023/1230 of the European Parliament and of the Council of 14 June 2023 on machinery \cite{europaEURLexx} foresees the use of autonomous systems for moving machinery and requires users and manufacturers to have adaptive systems depending on environment-related risks as well as requiring the presence of technical documentation on every constituent element of such systems. From a more general standpoint, the newly adopted Artificial Intelligence (AI) Act (Regulation (EU) 2024/1689) \cite{europaEURLexx} includes rules for AI systems being embedded into products, requiring full compliance of high-risk systems with the regulation.

Concerning the two latter texts, de Koning et al. provide us with a use case to evaluate the compliance of such a system while establishing the limitation of current standards in terms of AI, pointing out their inadequacy to comply with the regulation. They proposed an integrated risk management system that enables compliance with the concerned machinery regulation \cite{deKoning2024}.

As mentioned in Section \ref{sec:Background} the recently adopted Cyber Resilience Act (CRA) \cite{europaEURLexx} shows great advances in terms of software compliance. Indeed, the CRA requires the CE marking for specific software. Focusing more on Class II systems listed in the Annex III of the same document we can see that the EU understands the great importance and criticality of OT systems as they require a more thorough CE assessment. While the CRA adds pressure on software and system manufacturers to deliver quality, security and safety, the test requirements still need to be elaborated from legislative requirements. This critical step in the conformity process still needs to be addressed as only the technical documentation of the system needs to be assessed for the CE marking.

\section{Identification of challenges}
\begin{figure*}
    \centering
    \includegraphics[width=1 \textwidth]{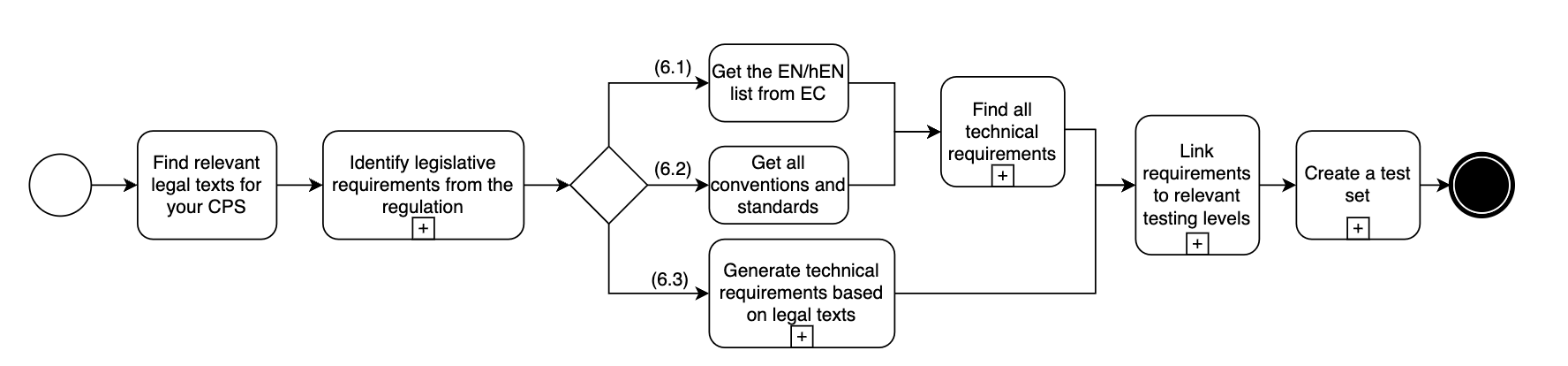}
    \caption{The exploratory path towards test generation based on a (EU) legal text}
    \label{fig:four}
\end{figure*}

Figure \ref{fig:four} shows the steps to explore the various paths starting from a single legal text. For each path, we give an example and discuss possibilities to automate the information gathering, leading to specifications and requirements that can be used to generate tests. In general, legal texts might be related to other legal texts, for example, implementing acts emanating from a Regulation or a Directive. Due to space constraints, we limit our discussion to legal texts found on the EUR-Lex platform \footnote{\url{https://eur-lex.europa.eu/}} based on keywords related to a chosen domain of application. 
For example, when looking at regulations related to cars used for people's transportation within the EU, we can search for \textit{Car} + \textit{Transport} + \textit{Vehicle} + \textit{People}. Doing so leads us to EU regulation 2019/2144 (as shown in Table \ref{tab:reg_stan_tab1}). However, this is only one regulation, and many more could be relevant when talking about producing or selling cars used for people's transportation within the EU. The EC offers the \textit{My Trade Assistant} tool on the \textit{Access2Markets} platform \footnote{\url{https://trade.ec.europa.eu/access-to-markets/en/home}} where one can get relevant information about a product. 
In this case, the product we look for falls under Section 17, Chapter 87 Heading 8703 over \textit{Motor cars and other motor vehicles principally designed for the transport of persons} following the Harmonised System Code of the World Customs Organization. The tool allows us to find the 3 framework regulations and links to relevant legal texts. We also find information related to \textit{Technical Harmonisation} specific to vehicles in the EU, allowing for any vehicle with a valid \textit{Certificate of Conformity} in an EU country to be marketed in any EU-wide country. By reading the first framework regulation, we find references to related ISO standards. We also observed experimental features on the \textit{EUR-Lex} platform \cite{europaEURLexx} providing us with a graph showing the relationship between a legal text and other material however, the only visual information is based on the color of a node (legal basis, citations, amendments, etc.) not even the name of related materials (the document number pops up when hovering on a node) and it doesn't include EU or international Standards.

\vspace{12pt}
\begin{quote}
    \textit{\textbf{Challenge 1: }Finding specific concerns from relevant legal texts based on keywords or product references.}
\end{quote}
\vspace{12pt}
For the first path, we will look into legal texts that were the subject of a harmonization request from the EC (3.1 in Figure \ref{fig:four}). From the EU eNorm platform,\footnote{\url{https://ec.europa.eu/growth/tools-databases/enorm/}} we find a relevant regulation with a mandate for a hEN request. 
Then, we assess whether the referenced legal text is a Directive or a Regulation and, if not, find it (i.e., in the case of an implementing or a delegated act). After that, we searched on the website of the relevant ESOs for the hEN related to the request. Then, we look into the hEN's content to see its specification level. Eventually, we see if it refers to other ENs and international standards or specifications. Finally, from the corpus of collected documents, we extract the lowest-level specifications that we could find and establish whether a technical test requirement can be created based on it. If not, we try to see what could be done to reach a sufficiently low specification level to make such test requirements and their corresponding tests.

For the legal texts that were not subject to a request from the EC to produce a hEN (3.2 in Figure \ref{fig:four}), we read them to search for direct references to specific international standards or conventions. In parallel, we also look into direct references to legal texts from EN on ESO websites. Indeed, during our research, we found that ESOs also referred to EN based on specific regulations. While it could be due to a desire from those organizations to help, we believe those links are artifacts remaining from previous but no longer active mandates. After selecting a legal text, we resume assessing the sufficiency level of the technical specification we found at the end.

Finally, for the regulations that do not mention any material that is or can lead to a specification (3.3 in Figure \ref{fig:four}), we look into solutions and research that have been carried out in various domains. We do not look into other standards and internal processes from specific companies. Although, this is a standard method to request a CE marking on a product, requiring the manufacturers to prove that their implementation respects relevant legal texts.

To summarise, we found three paths to go from a legal text to a technical specification:
\begin{enumerate}
    \item[(6.1)] Harmonised European standards (hEN) and European standards (EN)
    \item[(6.2)] International standards and conventions mentioned in the legal texts
    \item[(6.3)] Elaborating tests directly from the legal texts
\end{enumerate}

\subsection{Harmonised European standards (hEN) and European standards (EN)} \label{sec:firstpath}

The only difference between hEN and EN remains in the possibility to link standards to a specific EU regulation or directive, which is always possible with hENs while it is not always the case for ENs. As previously mentioned, hENs are meant to ease the approval of a CE marking. Those markings are primarily directed towards physical objects. However, they also mention interoperability, communication (such as radio devices), and software in the case of medical devices.

For example, we consider Regulation (EU) 2016/746 \cite{europaEURLex32017R0746} and Regulation (EU) 2016/745 \cite{europaEURLex32017R0745} on medical devices as previously cited in  Table \ref{tab:reg_stan_tab1}. There is a mandate (nr. \textit{575}) for those regulations on the EU eNorm Platform. By following the platform link to the register of commission documents \cite{europaRegisterCommission}, we get the collection of standards related to the mandate and can look for relevant ones. This list includes, among other things, the \textit{IEC EN 60601 standard on medical electrical equipment} \cite{iso606011112015} of which the cost of the first cited part (General requirements for basic safety and essential performance) is already CHF 1300 \footnote{https://webstore.iec.ch/en/publication/67497} (out of 55 citations).
\vspace{12pt}
\begin{quote}
    \textit{\textbf{Challenge 2: }Access to cited material is not direct nor free of cost.}
\end{quote}
\vspace{12pt}
The hEN also refers to \textit{EN 82304-1}, which establishes the difference between \textit{ISO/IEC/CEN 82304 - Health software general requirements for product safety} \cite{isoISOTS8230422021} categories of software products that fall into \textit{Regulation (EU) 2016/745} and those that do not. The second part of this standard \textit{Health and wellness apps - quality and reliability} refers in turn to \textit{ISO/IEC/IEEE 9945-2009, information technology, Portable Operating System Interface (POSIX), Base Specifications, Issue 7 - Taking into account Technical Corrigendum 1 and 2} \cite{IEEE5393893}. Here is the header of a method specification that can be found in POSIX \cite[p.~1816]{IEEE5393893}: 
\vspace{12pt}
\begin{quote}
    \textit{sem\_getvalue() - get the value of a semaphore ...}
\end{quote}
\vspace{12pt}
Furthermore, this base specification offers a deeper level of normative and informative references leading to communication protocol specifications such as Request For Comments (RFC) from the International Internet Task Force (IETF), e.g., \textit{RFC 791 on Internet Protocol Version 4 (IPv4)} \cite{rfceditor791Internet}. 

However, following such a path fails to keep the traceability between the different specifications and a specific article in the legal text. 
\vspace{12pt}
\begin{quote}
    \textit{\textbf{Challenge 3: }Identifying the percentage of compliance with the legal text reached after complying with a technical requirement is not straightforward.}
\end{quote}
\vspace{12pt}
\subsection{International standards and conventions mentioned in the legal texts}

By looking in the \textit{Document summary} of the \textit{Directive 2014/90/EU on marine equipment} \cite{europaEURLexx}. We can establish a relation to \textit{Commission Implementing Regulation (EU) 2022/1157 as regards design, construction, and performance requirements and testing standards for marine equipment}. However, this implementing regulation was repealed by \textit{Commission Implementing Regulation (EU) 2023/1667} which was also repealed by \textit{Commission Implementing Regulation (EU) 2024/1975} with the same regards \cite{europaEURLexx}. This implementing regulation includes all the standards required to satisfy the directive. It mostly refers to an international convention called \textit{Safety Of Life At Sea} (SOLAS) \cite{IMO1974}. The articles from that convention also point to standards from the International Maritime Organisation (IMO), IEC, and ISO alongside the ESO.

For example, point \textit{MED/5.22} on \textit{Ship Earth station for use in the GMDSS} (Global Maritime Distress and Safety System) refers to SOLAS \textit{CHAPTER IV. Radiotelegraphy and radiotelephony} article 10. This specific point refers to \textit{IEC 61097-16:2019} \cite{iec61097162019Webstore}. However, we have already explored a similar path, so we will look into \textit{ETSI ETS 300 460:1996 + A1:1997}, a specific sub-point for Inmarsat (brand) devices. This standard offers verbose specifications with a lot of possible interpretations. For example we have \textit{6.8 Operating controls}: 
\vspace{12pt}
\begin{quote}
    \textit{Purpose: Equipment shall be so constructed that it can be operated properly and readily by a suitably qualified member of a ship's staff.} 
\end{quote}
\vspace{12pt}
Such non-functional requirements could be relevant at multiple testing levels. In this case, it is not only \textit{Acceptance Testing} but overall \textit{Hardware Testing} in the context of the user interface without forgetting the \textit{Human} component.

\vspace{12pt}
\begin{quote}
    \textit{\textbf{Challenge 4: }Elaborating tests for all relevant testing levels from functional or non-functional requirements requires a broad expertise.}
\end{quote}
\vspace{12pt}

\subsection{Elaborating tests directly from the legal texts}

The primary purpose of this paper is to find a testable specification and the corresponding test requirements using official material and referenced standards while keeping the possibility of tracing the original legal text. Nevertheless, we could also try to extrapolate features related to software engineering (or general IT and OT concerns) from original legal texts. 
\vspace{12pt}
\begin{quote}
    \textit{\textbf{Challenge 5: }Determining technical requirements from legal texts remains ambiguous.}
\end{quote}
\vspace{12pt}

\section{Discussion}
We have cycled through the different possible processes to produce a legally compliant CPS. While the EU has already put great effort into helping companies comply with legislative requirements (i.e., EN and hEN), challenges remain. Nevertheless, various methods and technologies laid out in Section \ref{sec:RelatedWork} present an encouraging approach that would enable automatization in the compliance checking process. In this section, we discuss which technologies could help solve each challenge.

For the first challenge on \textbf{finding specific concerns from relevant legal texts based on keywords or product references}, the \textit{My Trade Assistant} tool of the \textit{Access2Markets} platform provides valuable information related to products. When looking at the different documents provided and related materials, we can manage to find many details. Most certainly, all the information is neither necessary for all associated products nor relevant when looking at compliance with a specific CPS. As mentioned in Section \ref{sec:RelatedWork}, existing work on requirements engineering using NLP or LLM technologies could help extract relevant features and related articles from legal texts. Then, the \textit{My Trade Assistant} tool could be embellished with a feature-based search engine, allowing the precise identification of a CPS (or any product) and producing the relevant articles. Examples of existing feature-based tools include the AI act compliance checker provided by the Future of Life Institute \cite{artificialintelligenceactComplianceChecker}, which already helps actors determine the applicability of certain legal texts in specific contexts.

Concerning the second challenge as \textbf{access to cited material is not free of cost} we find it weird that even harmonized standards with the sole purpose of helping manufacturers comply with a mandatory legal text are not free. More, most often, they do not directly include testable system specifications. Standards mentioned in legal texts or required by the EC should be free for any interested party (consumers for market surveillance, manufacturers for building compliant products, etc.). Indeed, the intention is not to lose people in complexity and send fines afterward (e.g., when an incident shows a failure to comply) but rather to provide EU citizens with safe and secure products. Interestingly, in France, for example, the NSB (AFNOR) offers mandatory norms for free following a national decree (\textit{décret n° 2021-1473 du 10 novembre 2021 (Art.17)}). Nevertheless, we will soon be able to consider this challenge as solved following the ``\textit{Malamud}'' case from 5th of March 2024 (C-588/21 P) where the European Court of Justice decided that the EU will have to offer hEN for free \cite{eurogipFreeAccess}.

The third challenge on \textbf{identifying the level of compliance with the legal text reached after complying with related specifications}, we can get inspiration from Bicaku et al. as explained in Section \ref{sec:RelatedWork} \cite{Bicaku2019}. They built a working framework for assessing compliance with standards using technology-specific tests. Their framework is extremely powerful in terms of compliance checking while keeping the percentage of compliance as each \textit{control} is linked to a specific point in the refereed standard. However, such a method might require work from many (if not all) actors of the various domains of application as the framework requires each technical requirement to have a technology-specific control (i.e., the \textit{Managing Agents} in Fig. \ref{fig:four}) while the MIP would remain quite similar across technologies. The model-based approach from Sacré et al. also offers great capabilities to keep track of the degree of compliance of a CPS. Indeed, the \textit{distance} between the tested model and the legally compliant model could be used as an indicator of compliance in terms of what is still required from the tested model to become fully compliant \cite{Sacre2021}.

Looking at the fourth challenge pointing out the complexity when \textbf{elaborating tests for all relevant testing levels from functional or non-functional requirements} we can look back at Section \ref{sec:RelatedWork} on cross-levels testing methods suggested by Huang et al. \cite{Huang2010}. However, their method implies a configurable middleware connected to the System Under Test (SUT) and would need to be adapted every time it is used to test a different system. Furthermore, it does not include the possibility to test the \textit{Human} component of a CPS. Considering \textit{Network Testing} and the communication protocol specifications we found when following the first path in Sec. \ref{sec:firstpath}, research in the cyber-security community might be used for automatisation. Indeed, we found works aiming at attacking an unmodified implementation of a transport protocol based on its representation as a Finite State Machine (FSM) \cite{Jero2015} and even suggesting ways of generating said FSM from the verbose specifications as it was the most time-demanding manual step \cite{Pacheco2022-au}. While their goal was to find vulnerabilities in communication protocol to leverage attacks on all systems implementing those, we can still leverage their work for testing a correct protocol implementation based on an FSM \cite{EIFakih2004}. Other methods exist to test OT communication protocols, such as using test beds developed specifically for this purpose \cite{Boeding2023}. However, such methods require extensive work to test only one communication protocol and might not be the preferred way to deploy a greater-scale test methodology (as with Huang et al. \cite{Huang2010}. In the case of CPS and considering the type of information we have at hand, fuzzing \cite{Eisele2022} and model-based testing \cite{Aerts2017} seem to be the most relevant testing techniques for larger scale and automatized deployment. However, even if we could go from such specifications to the generation of tests, it requires other tools to automatize the process fully.

Finally, the last challenge of \textbf{understanding technical requirements from legal texts} is certainly the most challenging one. Indeed, legal texts are not built like technical specifications, legal texts are legal instruments protecting EU citizens. However, legislators need to understand the specificity of the different markets and related use cases to build such texts. Interested parties would greatly benefit from understanding what is expected of them directly. Nevertheless, encouraging work allows for the automatized retrieval of requirements from legal text as explained in Section \ref{sec:RelatedWork}. Indeed, this challenge could be related to the first one by using the \textit{My Trade Assistant} tool of the \textit{Access2Markets} platform to produce not only relevant articles but relevant legislative and technical requirements from both legal texts and related standards. Furthermore, intuitively we could think that when building legal texts, legislators need to lay down such requirements beforehand to produce a text with market relevance. Those requirements could be shared with manufacturers at the same time as the legal text to help manufacturers understand more precisely what is expected of them. If it's not already the case then legal text should definitely be empirically informed. Indeed, even though initial analyses have been thoroughly performed for the creation of original legal texts, they should be adapted, taking into account the fast-changing technology landscape and needs of the various actors \cite{Sunstein2011-js}.

\section{Conclusion and Future work}
On one hand, we showed the overall context and the challenges related to testing the compliance of CPS on the EU market. While we did not solve those challenges, we found existing and encouraging research one could leverage for solving some of them. However, the use of automated tools such as NLP and LLM could be problematic when considering the lawfulness of the results produced (features, requirements, applicable articles, etc.). On the other hand, we saw that while EU institutions were providing market actors with a tremendous amount of resources, it might be complex to understand what is expected of manufacturers (and what citizens could expect from them). Besides that, there are not many (positive and helpful) incentives for the manufacturers to comply with EU legal texts. Indeed, we showed that the CE marking could be self-assessed while the EU only foresaw deterrents in the shape of minimal fines directly added in EU Directives or left to member states in case of failure to comply using the \textit{Market Surveillance} mechanism.

Legal texts are driving the way products are made and delivered to the market. While they are not sufficiently technical for engineers and manufacturers to understand what is expected from them, they add constraints on the final implementation. However, in some cases it is possible to find sufficiently technical specifications on which one could base compliance tests.  While we managed to find related regulations for each domain, they certainly do not cover all the various CPS that currently exist or will be created in the future. However, the newly adopted CRA includes concerns about hardware and software in general thus CPS). Nevertheless, this legal text remains high level in terms of technical requirements and still calls for the creation of relevant hEN. It might be interesting to follow the extensive work that has been performed on regulations for medical devices in order to identify exactly the current legislation and related industry standards for each application domain. The EC should support such initiatives as well as suggest approved technology. Indeed, when going from legislation to a (testable) specification, we often need to buy in-between standards, even when ending up on access-free specifications (i.e., IETF RFC). While it should no longer be the case in the future following the \textit{Malamud} case it is still the case at the moment. This also makes research more complex. It would greatly improve the overall process if the suggested path between the development and the approval of a product could be officially laid out before starting a project before consulting industry experts directly. This includes holistic, relevant, and related legal texts, technical requirements from those texts, as well as related standards (e.g., a centrally served project configuration tool).

Furthermore, before the newly adopted AI act, the CE marking was mostly (but not exclusively) related to physical objects. Fortunately, with the CRA, the EC now considers such marking for software and hardware products including CPS. While we wait for tools and materials enabling the technical implementation of the CRA, a temporary solution could be the application to each domain of an initiative similar to \textit{European Defence Standards Reference System} (EDSTAR), which is a collection of international standards selected and shared across defence actors. 

Finally, encouraging (and tremendous) work has been performed by the EU to ensure the safety of EU citizens. However, while some legal texts include the necessity for a revision after some time, it seems that the EU still lacks empirical information on the regulated market to eventually identify the decoupling between the intention behind produced legal texts and their practical implementation in the various domains of application. The only empirical indicator lies within the \textit{Market Surveillance} mechanism and the presence (or non-presence) of a product that failed to comply but this indicator might be biased as the \textit{Market Surveillance Authorities} might not have the means to assess each product and might heavily rely on notification from citizens or other actors after a failure to comply led to a damage of any kind. The CRA would also include the monitoring of software vulnerabilities found after a product was sent to the market as a mean to evaluate the efficiency of the measures. As CPS is becoming the standard (i.e., ADAS, Autonomous Machinery, and future prospects on embedded AI) the safety of EU citizens will heavily depend on the quality of the (mandatory) tests before the introduction on the market as a failure to comply might lead to human casualties before correction via the \textit{Market Surveillance} mechanism. 

\bibliographystyle{IEEEtran}
\bibliography{IEEEabrv,99.biblio}

\appendices

\section{Data Availability}
The data used in this paper is the official material from the European Commission, links are available either in the citations or in the footnotes of the paper. 

\section{Contribution of the authors}
G. Nguyen is a researcher and PhD student with qualifications in business engineering and computer science as well as professional experience in terms of compliance checking with Cybersecurity (technical and privacy) related concerns in European legal texts. This author used available material from the European Union (European Commission, more specifically) to analyze and understand the European context when checking the compliance of CPS.

M. Knockaert is a researcher with qualifications and expertise in European legislation and reviewed first author's understanding and writings on EU-related points at earlier stage of this paper.

M. Lognoul is a researcher with qualifications and expertise in European policies and reviewed first author's understanding and writings on EU-related points in this paper.

X. Devroey is a professor of computer science with extensive expertise in software testing. This author is also the supervisor of G. Nguyen and reviewed the whole paper for quality and content.

\end{document}